# Ballast water-mediated species spread risk dynamics and policy implications to reduce the invasion risk to the Mediterranean Sea

*Zhaojun Wang[1,*], Mandana Saebi[2], Erin K. Grey[3], James J. Corbett[1]*

[1] *University of Delaware, 305 Robinson Hall, Newark, DE 19716, USA*

[2] *University of Notre Dame, 384 Nieuwland, Notre Dame, IN 46556, USA*

[3] *University of Maine, 172 Hitchner Hall, Orono, ME 04469, USA*

*\*Corresponding author: izhaojun@udel.edu*

**Abstract**

The Mediterranean Sea is one of the most heavily invaded marine regions. This work focuses on the dynamics and potential policy options for ballast water-mediated nonindigenous species to the Mediterranean. Specifically, we (1) estimated port risks in years 2012, 2015, and 2018, (2) identified hub ports that connect many clusters, and (3) evaluated four regulatory scenarios. The risk results show that Gibraltar, Suez, and Istanbul remained high-risk ports from 2012-2018, and they served as hub ports that connected several spread clusters. With policy scenario analysis, we found that regulating the high-risk hub ports can disproportionately reduce the overall risk to the Mediterranean: the average risk to all ports was reduced by 5-10% by regulating one high-risk hub port, while the average risk to all ports was only reduced by 0.2% by regulating one average-risk Mediterranean port. We also found that only regulating high-risk ports cannot reduce their risks effectively.

**Keywords:** Ballast water management, Mediterranean, marine invasion, global shipping, policy implication

## 1. Introduction

The Mediterranean Sea is a biodiversity hotspot that is heavily impacted by a variety of human activities, including overfishing, eutrophication, and the introduction of nonindigenous species (NIS) (Kleitou et al., 2021). Although it is widely believed that the majority of new NIS introductions fail, some do succeed in establishing populations in new areas where they can become "invasive" by causing harm to ecosystems, economies, or human health. The rapid spread and conspicuous impacts of a pair of invasive chlorophytes helped raise awareness of the problem NIS in the Mediterranean (Meinesz et al. 2002; Verlaque et al. 2004; Klein & Verlaque 2008). To date, nearly 1000 marine species and over 660 multicellular species are recognized as nonindigenous to the Mediterranean Sea, even though the recorded total may be underestimated (Galil, 2012, Zenetos et al., 2012, Zenetos et al., 2010, Bariche et al., 2015). Efforts are needed for scientifically robust, sensible, and pragmatic plans to minimize introductions of marine NIS and to study those present (Galil et al., 2014).

Due to the logistical and management difficulties in targeting specific species, vector management is a more efficient way to control biological invasions (Carlton, 2003). Identifying the major introduction vectors is a first step to prioritize vector management efforts with the aim to reduce new species arrivals. Of the NIS recorded in the Mediterranean Sea, 22% are from vessels, 14% are considered to have entered through the Suez Canal and secondarily spread by vessels, and 4% are introduced by mariculture and secondarily spread by vessels (Galil, 2012). Due to the geographical location, the Mediterranean is crisscrossed by busy shipping traffic, making shipping a significant vector in the primary and secondary dispersal of 185 alien species (Galil, 2012). Katsanevaskis et al. show over 50% of these species were introduced by shipping (Katsanevakis et al., 2013).

The high NIS introduction rate from shipping activities is due to the busy shipping traffic to the Mediterranean. For example, the Mediterranean region experienced significant growth in the number of vessels calling between 2011 and 2016 (from 10% in Spain to 20% in Croatia), compared to the decreases in Northern ports (Viana et al., 2020). There are concerns that the environmental issues face more challenges due to potential increasing traffic, given the general increase in international trade, the role of the Mediterranean in the Twenty-First Century Maritime Silk Road Initiative (Chaziza, 2018), and the expansion of the Suez Canal in 2015.

In the context of ballast water-mediated harmful aquatic organisms and pathogens (HAOP), the 2004 Ballast Water Management (BWM) Convention at the International Maritime Organization (IMO) is an international regulation that is uniformly applied across the commercial fleet. No specific attention is paid to local areas at the IMO despite the disparate NIS spread risk globally (Saebi et al., 2020b). In this regard, this work aims to inform control ship-borne invasions to the Mediterranean. There are two main ship-borne vectors, ballast water discharge and biofouling. This work focuses on ballast water-mediated species introduction because models and policy for this vector are more developed.

Our modeling work builds off of previous studies that have attempted to quantify the invasion risk caused by ballast water discharge. The first global biological ship-born NIS spread models were based on ship trajectories (Drake and Lodge, 2004, Kaluza et al., 2010, Keller et al., 2011), but did not consider environmental factors that are known to be important in determining whether NIS establish or not (Seebens et al., 2013). Seebens et al. further consider survival and establishment pressures of NIS based on ship trajectories and environmental factors (Seebens et al., 2013). This model has been used to predict invasion risks (Seebens et al., 2016, Sardain et al., 2019); however, it implicitly assumes that the invasion risk to a port only comes from the last port of call. Xu et al. further construct a higher-order network based on the first-order invasion risk model of Seebens et al. to capture the invasion risk including several previous ports of call (Xu et al., 2016, Saebi et al., 2020b). Accordingly, we use the model of Xu et al. and Saebi et al. to estimate the NIS spread risk to the Mediterranean. We also use cluster analysis to identify the high-risk ports that connect different clusters.

Based on the risk assessment and network analysis results, we designed four policy scenarios and examined two regulatory standards to reveal the effective ways to reduce species spread risk to the Mediterranean. We found different ways for the high-risk hub ports and for the Mediterranean as a whole.

To explore the potential impact of the Suez Canal expansion, shipping traffic, and ballast discharge dynamics on NIS spread risk, we analyzed shipping data from 2013, 2015, 2018. This allowed us to observe changes in shipping traffic and resultant ballast NIS spread post-Suez expansion (in 2015). While many feared that the canal expansion would increase NIS introduction risk to the Mediterranean, previous studies found that the number of containerships and related vectors actually decreased (Shenkar and Rosen, 2018, Galil et al. 2015).

## 2. Methods and data

### 2.1 Data

Global shipping movement records are from Lloyd's List Intelligence (LLI) of three years: 2012, 2015, and 2018 (starting on May 1$^{st}$ of these years and ending on April 30$^{th}$ of the following years). Each record of the dataset is composed of origin and destination ports, departure and arrival time, vessel identifier, and vessel specifications (deadweight tonnage, flag, year of built, length, width, and draft). Though the study is focused on Mediterranean ports, it is necessary to consider all shipping movements worldwide when calculating species spread risk. This is because we are using the higher-order network to consider the full trajectory of vessel movement, and ballast water discharged at ports outside the Mediterranean may generate risk to the Mediterranean via the higher-order network.

Ballast water discharge volume and frequency are estimated with data from US ports from the National Ballast Information Clearinghouse. The dataset includes vessel IMO number, ballast water discharge volume, and vessel type. We used records for the years 2004 to 2016 and a random forest method to estimate the ballast water discharge volume of each vessel movement (Saebi et al., 2020a). Records with missing information are removed.

Environmental data (temperature and salinity) of ports are from the Global Ports Database (Keller et al., 2011) and the World Ocean Atlas (Zweng et al., 2013, Locarnini et al., 2013). Ecoregion information is from Marine Ecoregion of the World (Spalding et al., 2007) and Freshwater Ecoregion of World (Abell et al., 2008).

## 2.2 Higher-order species flow model

The Species Flow Higher-Order-Network (SF-HON) has been used to estimate the species spread probability considering the higher-order pattern (Saebi et al., 2020b, Saebi et al., 2020a). The SF-HON model performs better in species spread risk estimation by capturing a more complete ballast water discharge profile given that ships may not discharge all ballast water at one port and usually discharge the remaining ballast water to several calling ports (Saebi et al., 2020b). The SF-HON model integrates shipping moves, vessel specifications, biogeographical information of ports, and temperature and salinity similarity. Similar to other methods to model ship-borne species spread risk (Seebens et al., 2013, Sardain et al., 2019, Drake and Lodge, 2004), the SF-HON model adopts a stage-based framework widely used in invasion biology, whereby nonindigenous probability, species transport probability, and establishment probability. See Equation 1, which estimates the species spread relative risk from port $i$ to port $j$ of vessel $v$.

$$p(NIS\ spread)_{ij}^v = p(nonindigenous)_{ij} \times p(intro)_{ij}^v \times p(establish)_{ij} \qquad (1)$$

The nonindigenous probability is 0 if both origin and destination ports belong to the same or neighboring ecoregions, and 1 otherwise, since neighboring ecosystems are likely to share more species due to natural dispersal (Costello et al., 2017, Abell et al., 2008). The probability of being nonindigenous is included because we assume that species are introduced to novel regions by ships and not by natural vectors.

$$p(intro)_{ij}^v = \rho^v (1 - e^{-\lambda D_{ij}^v}) e^{-\mu \Delta t_{ij}^v} \qquad (2)$$

Equation 2 describes the transportation probability, where $D_{ij}^v$ is the amount of ballast water discharged at a port; $\lambda = 3.22 \times 10^{-6}$, is the species introduction potential per volume of discharge (Seebens et al., 2013); $\mu = 0.02$, is the daily mortality rate of species in ballast water during the voyage, $\Delta t_{ij}$ is the duration of one shipping voyage; $\rho^v$ measures ballast water treatment efficacies.

$$p(establishment)_{ij} = \alpha e^{-\frac{1}{2}\left[\left(\frac{\Delta T_{ij}}{\delta T}\right)^2 + \left(\frac{\Delta S_{ij}}{\delta S}\right)^2\right]} \qquad (3)$$

Equation 3 is the establishment probability, which depends on environmental similarity (temperature and salinity) of the origin and destination ports. $\Delta T_{ij}\ and\ \Delta S_{ij}$ are the temperature and salinity difference of the origin and destination ports; $\delta T$ and $\delta S$ are the standard deviations in temperature and salinity, respectively, where $\delta T = 2\ °C$ and $\delta S = 10ppt$.

$$p(NIS\ spread)_{ij} = 1 - \Pi_{r,v}(1 - p(NIS\ spread)_{ij}^v) \qquad (4)$$

Equation 4 is to aggregate over all routes *r* of a ship *v* from port i to port j, to get the risk of each port pair. The species spread risk obtained from Equation 4 is first-order risk, which will be the input to the SF-HON algorithm (Xu et al., 2016) to calculate HON risks.

The HON algorithm generates HON of species flow and builds the physical adjacency matrix in which each edge weight is calculated by averaging over all HON edges corresponding to that port pair. Finally, the cumulative risk of each port *k* is the aggregation of the NIS spread risk over all incoming ports with Equation 5.

$$p(NIS\ spread)_k = 1 - \Pi_i(1 - p(NIS\ spread)_{ik}) \qquad (5)$$

In this work, the spread risks are calculated based on the global shipping traffic and filtered to focus on Mediterranean Sea ports. Port risks of years 2015 and 2018 are normalized with regard to the highest port risk in 2012 to examine the risk change over time.

## 2.3 Cluster analysis

Cluster analysis provides a large-scale view of ship-borne species spread risk and identifies the groups of ports within which ports are more connected and more likely to share species (Saebi et al., 2020b). Such information can be useful for policymakers who wish to most effectively reduce global HAOP spread. The main goal of *Infomap* is to find groups of nodes among which the species flow is quick and easy, which can be aggregated as a separate cluster. *Infomap* identifies clusters by optimizing the entropy corresponding to intra-cluster and inter-clusters using a recursive random-walk method. This method is suitable for extracting modules of species flow since random walks are the most similar to the species flow pattern.

It is important to note that we perform clustering on the higher-order network. In the HON network, several nodes may correspond to one port in the real world. For example, all higher-order nodes A|B, A|B.C, and A|D.C.E corresponds to port A. Given that, the number of nodes is an indicator of the number of higher-order patterns in the underlying data. The higher the number of nodes in the HON, the higher is the number of higher-order dependencies (Saebi et al., 2020b). Since multiple higher-order nodes may correspond to a single port, it is possible that a port belongs to multiple clusters.

## 2.4 Ballast water regulatory scenarios toward the Mediterranean

From the above analysis, Gibraltar, Suez, Istanbul, and Algeciras were found to be high-risk ports in 2012, 2015, 2018. These ports also act as hubs that connect the most clusters, serving as bridges for easier species spread among different clusters. We designed four policy scenarios to examine if regulation targeting these key ports may generate disproportionate risk reduction compared to regulating all Mediterranean ports.

In the first, second, and third scenario, we choose the top one (Gibraltar), two (Gibraltar and Istanbul), and three (Gibraltar, Istanbul, and Suez Canal) strait ports with the highest risk as the regulating targets since their importance in the clustering pattern remain important in three years, making policy analysis more robust. The targeted high-risk ports are regulated by the current Ballast Water Management System (BWMS) and other Mediterranean ports and outside-Mediterranean ports are not regulated. The single treatment efficacy of current BWMS is about 76% and the efficacy is about 99% when the IMO Convention is fully met (Wang et al., 2021). In the fourth scenario, we regulate discharged ballast water to all Mediterranean ports (347 ports regulated) with the same treatment efficacy of 76% and 99%. The policy scenario changes the parameter $\rho^v$ in Equation (2), i.e., 1-$\rho^v$=76% or 99%.

We then compare the risk changes of Mediterranean ports under these four scenarios with the scenario when no port in the world is regulated to see the effectiveness of these four policy scenarios. The policy scenario examination is done based on data of 2018.

## 3. Results

We show three aspects of the risk analysis results: (1) risk, shipping traffic, and ballast water discharge volumes by Mediterranean country (Figures 1, 2); (2) risk, shipping traffic, and ballast water discharge volumes to individual Mediterranean ports (Figures 3, 4); (3) NIS spread risk origin in the high-risk Mediterranean ports and countries.

### 3.1 Mediterranean countries: arriving traffic, BW discharge, average risk

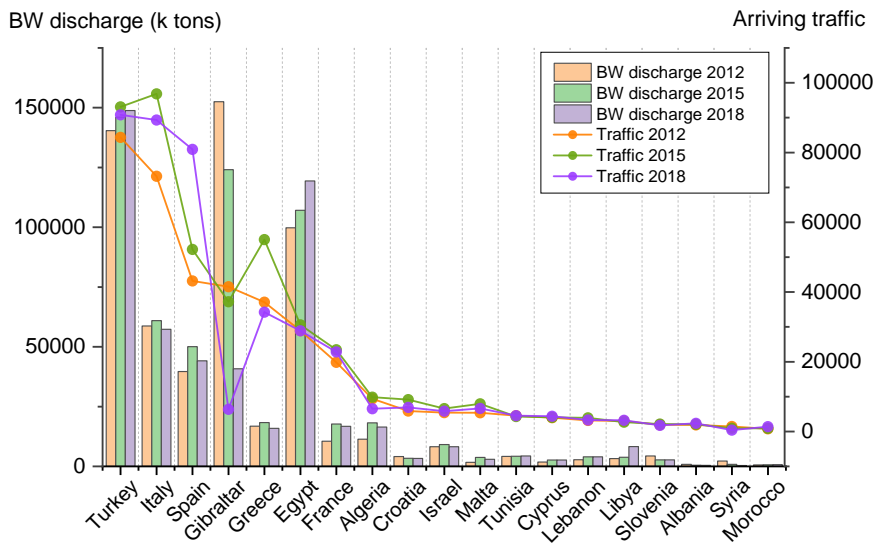

Figure 1 The arriving traffic and BW discharge to Mediterranean countries in three years (we list Gibraltar separately as itself). Ranked by arriving shipping traffic in 2012.

Only the arriving shipping traffic and BW discharge of Gibraltar experienced a sharp decrease from 2012 to 2018. The traffic to Spain increased a lot, so some traffic of Gibraltar may have diverted to Spain's ports given their close geographic locations. Egypt's traffic has no obvious change before and after the expansion of the Suez Canal, while the ballast water discharge to Egypt increased by 30%. This can be explained by increased vessel size and changed vessel type through the Suez Canal. Larger vessels tend to discharge more ballast water and bulk carriers and tankers tend to discharge more ballast water than container ships. This is proved by the traffic analysis via Suez Canal. From 2011 to 2018, there is a 2% increase in traffic in the Suez Canal and a 22% increase in total tonnage (Bereza et al., 2020). Also, the number of bulk carriers and tankers through the Suez Canal increased substantially while the number of containerships decreased a lot (Bereza et al., 2020).

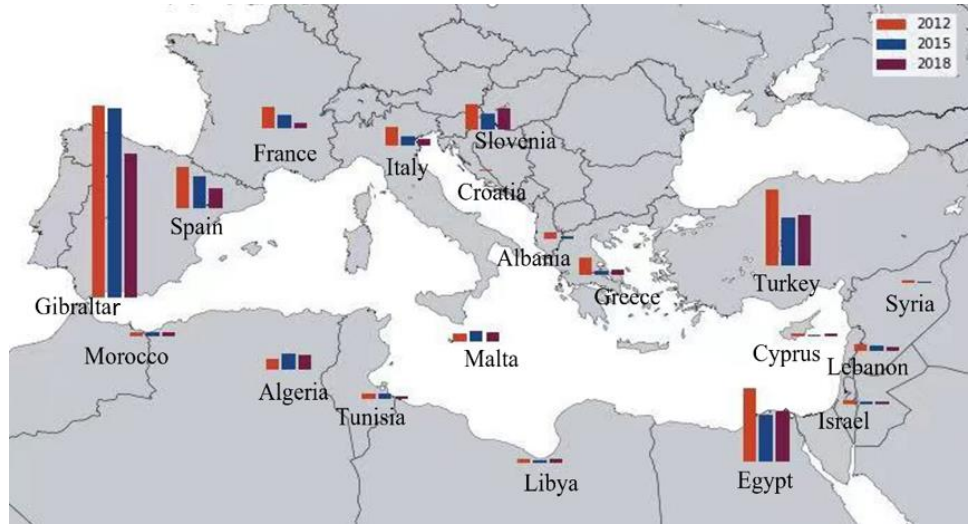

Figure 2 Average species spread risk of Mediterranean countries in each of three years

Figure 2 shows that the top three Mediterranean countries with the highest risks remain Gibraltar, Turkey, and Egypt in 2012, 2015, and 2018. No linear relationship between countries' risk change and traffic or BW change, except Gibraltar, whose risk change is positively related to its traffic and BW discharge volume. For example, more BW is discharged in Egypt and Spain, but the average port risks decrease over time. This can be explained by our higher-order risk assessment model. Table 1 shows the risk decrease is consistent with HON statistics. The HON pattern became less complex, shown by fewer HON nodes and edges in the risk spread network in 2018, compared to 2012. This shows the higher-order risk spread pattern plays an important role in determining the risk besides shipping traffic and BW discharge volume.

Table 1 Statistics of the higher-order pattern-Introduction & Dispersal to and within Mediterranean

|                     | 2012  | 2015  | 2018  |
|---------------------|-------|-------|-------|
| Number of HON nodes | 8553  | 6476  | 4879  |
| Number of edges     | 27313 | 24157 | 20026 |

## 3.2 Individual Mediterranean ports: arriving traffic, BW discharge, average risk and connecting clusters

At the port level, the traffic of Gibraltar of Istanbul in significantly higher than other ports in 2012 and 2015, but only Istanbul still has very high traffic volume in 2018 and the traffic to Gibraltar dropped dramatically in 2018. See Figure 3. At the same time, the traffic to most other ports increased. For example, a portion of the traffic increase to Algeciras may from Gibraltar, considering the close geographical location. Also, the traffic to Suez and Port Said increased, which is consistent with the fact of the expansion of Suez Canal in 2015 enabling more traffic going through Suez Canal instead of Gibraltar to the Mediterranean. The traffic diversion to the Suez Canal also explains the traffic decrease in Gibraltar.

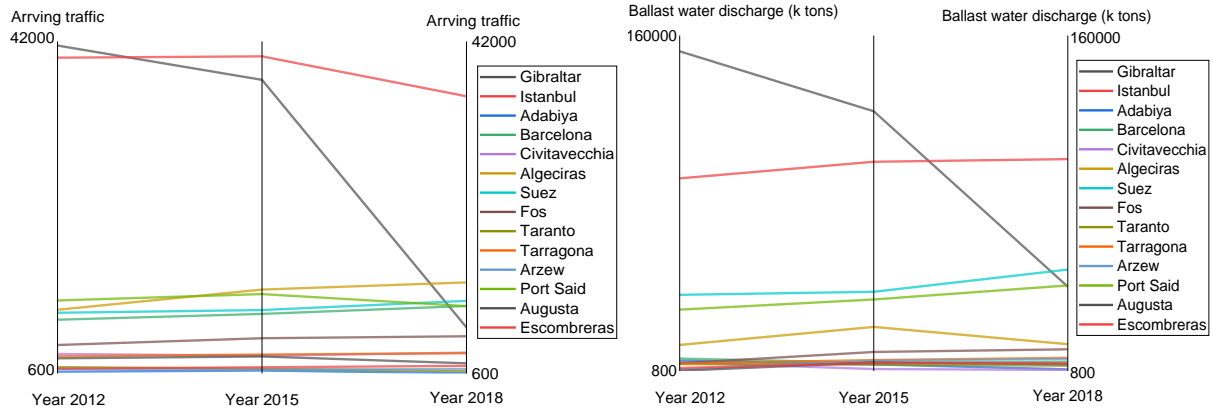

Figure 3 Arriving traffic and BW discharge changes in major high-risk Mediterranean ports

Figure 4 visualizes the locations of the top 14 Mediterranean ports with the highest risk in 2012, together with the risk changes over time. It shows that most high-risk ports experienced a risk decrease from 2012 to 2018, except Escombreras, Arzew, and Augusta. This risk decrease pattern can also be observed in Figure 4, where we plot the risk and the number of connected clusters of the top ten high-risk ports. In 2012, the risks of the top ten high-risk ports were all higher than 0.4, while the risks of nine high-risk ports were between 0.3 and 0.6 in 2015, and the higher bound of risks in 2018 further decreased to 0.8. Gibraltar is the port with the highest NIS risk in all three years, and its risk decreased by 25% from 2012 to 2018. Suez is another port with high spread risk and its risk decreased by 41% in 2015 and 45% in 2018, compared to 2012. The full list of port risks in three years can be found in Data availability.

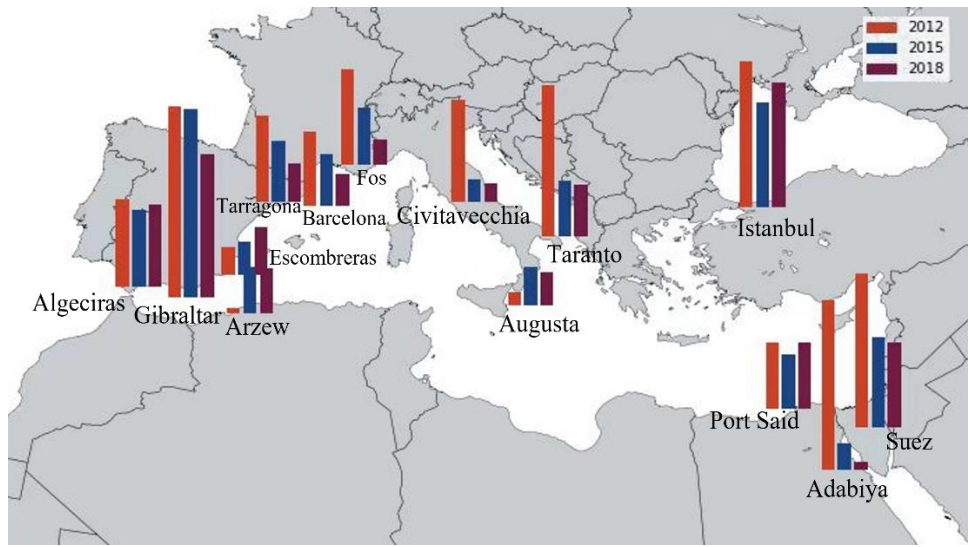

Figure 3 Species spread risk of high-risk Mediterranean ports in three years

We also observe the changes in the top ten high-risk ports. Arzew (in Algeria) and Port Said (close to Suez) become the top ten high-risk ports from 2015, replacing Adabiya (close to Suez) and Civitavecchia (in Italy). Then Augusta (in Italy) and Escombreras (in Spain) replace Barcelona (in Spain) and Fos (in France) in 2018 in the list. Despite the changes, Gibraltar, Suez, Istanbul, and Algeciras remain the high-risk ports over time. These four ports are "strait ports", connecting at least two different seas. Suez port is located at the Suez Canal, as one of the main hubs of global maritime trade and the main transit point between the Indian Ocean, the Red Sea, and the Mediterranean (Chaziza, 2018). Istanbul (in Turkey) connects the Mediterranean Sea and the Black Sea. Gibraltar and Algeciras are located at Gibraltar Straits, linking the Atlantic Ocean directly to the Mediterranean Sea.

By Figures 3 and 4, we can see BW discharge volume or shipping traffic are not the determinants in port risks, either, at the port level. Most ports experienced increased shipping traffic and BW discharge, yet decreased risks, except for Gibraltar and Istanbul. Please noted that the environmental conditions (i.e., temperature and salinity) modeled in our HON risk spread model are the same, so the decoupled patterns of port risks and BW volume cannot be explained by environmental parameters. We think two reasons may help understand this finding. (1) Gibraltar has a determinant role in deciding species spread risks of other Mediterranean ports. See Figures 3 and 4, where the traffic and BW discharge changes of Gibraltar experienced a substantial drop in 2018. (2) The higher-order network plays an important role in decreasing the risk. This is proved by the fewer HON edges and nodes in Table 1, as discussed in risks and traffic in Mediterranean countries.

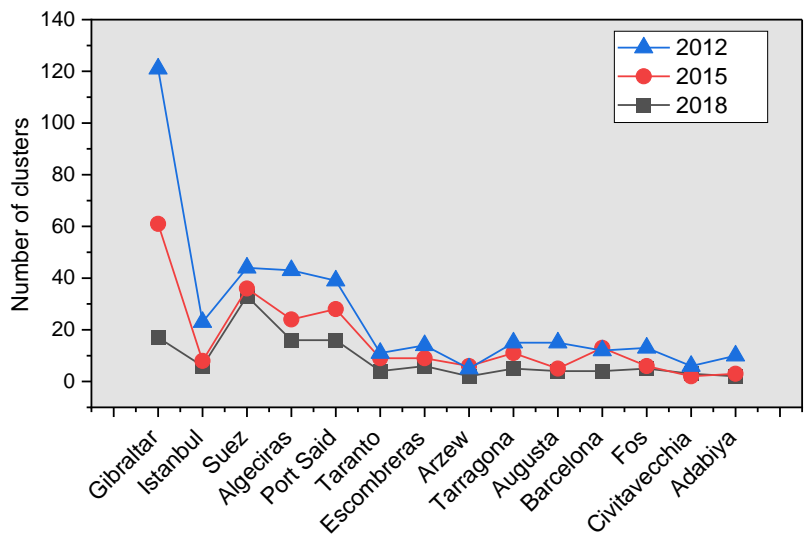

Figure 5 The number of clusters of high-risk ports in 2012, 2015, and 2018. Ports are ranked by the risk.

Figure 5 shows the changes in the number of clusters connected by high-risk ports. Ports within the same cluster have a higher possibility of species spread risks and, vice versa, ports in different clusters should have lower risks (Saebi et al., 2020b). However, ports, especially high-risk ones, connecting different clusters serve as "bridges" to make species spread more easily among different clusters, and therefore increasing the spread risk of the whole shipping work. Figure 4 shows that the number of clusters connected by these high-risk ports becomes fewer over time. This change is most obvious for Gibraltar with the number of clusters dropping from 121 in 2012 to 17 in 2018. Suez became the port connecting most different clusters, followed by Gibraltar, Port Said, and Algeciras. These four ports belong to many clusters because they are the spokes connecting the Mediterranean Sea with other areas. Clusters of all Mediterranean ports can be found in Data availability.

### 3.3 NIS spread risk origin

In this section, we examine the NIS sources to the Mediterranean. Our HON spread risk model reveals that the highest species spread risk to the Mediterranean are from inside the Mediterranean. See right panels of Figure 6. This is because of more traffic from this area. Med impact on Med becomes smaller (from 90% in 2012 to 80% in 2018). Western Africa and the Black sea are the major species risk sources besides the Mediterranean itself. These patterns are similar to individual high-risk ports (left panels of Figure 6).

Western Asia became more important in 2015 and 2018, no matter to the Mediterranean as a whole or to individual ports. For example, the biggest sources of species spread risk to Port Said are the Mediterranean and Western Asia in 2015 and 2018. The impact of Western Asia on Port Said became larger in 2018 and the impact of the Mediterranean became smaller. Given the risks of Port Said in the studied years did not change a lot (Figure 3), the reason is very likely to be the expansion of the Suez Canal in 2015, which makes more traffic coming to Port Said from the canal rather than the Mediterranean. (The reason for the emergence of Port Said in the left panels of Figure 6 is the risk decreases of other high-risk ports, which makes Port Said become one of the top ten high-risk ports in 2015 and 2018)

The species spread risk to Istanbul, another strait port most affected by the Mediterranean, Black sea, and Northern Europe, however, experienced a larger impact from the Mediterranean and a smaller impact from Northern Europe.

Despite the changes of proportions from different species origin areas, the major origin areas to the Mediterranean are those closer to the Mediterranean geographically. This can be explained by our HON risk model. The geographic distance is an important parameter in determining the introduction likelihood (Equation 2). The shorter distance means less duration of species transported in the ballast water tanks and a higher probability to be alive when the ballast water is discharged (Miller and Ruiz, 2014).

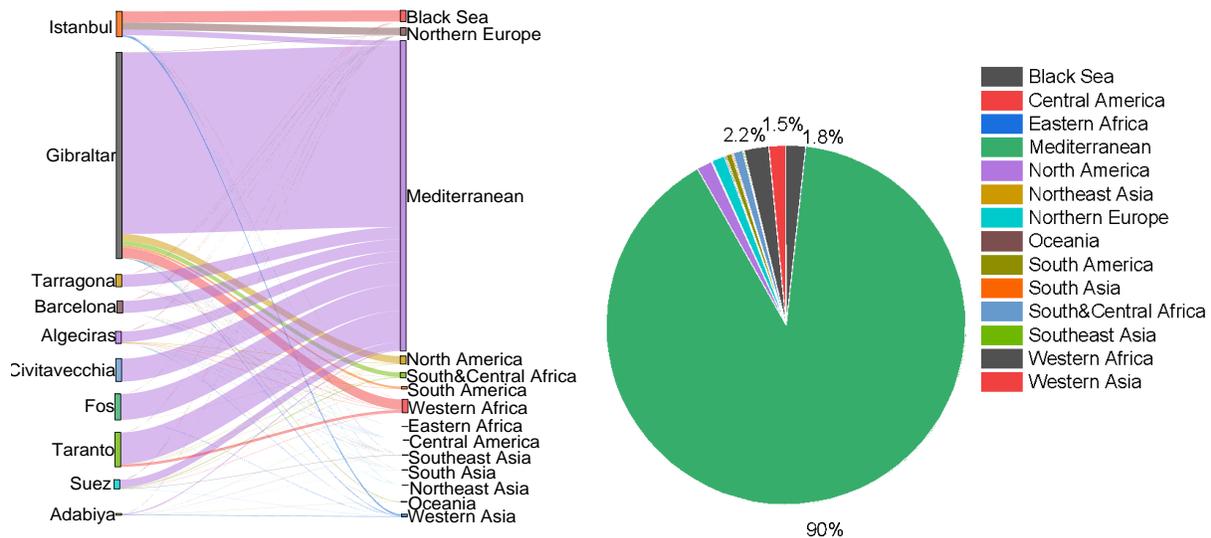

a. 2012

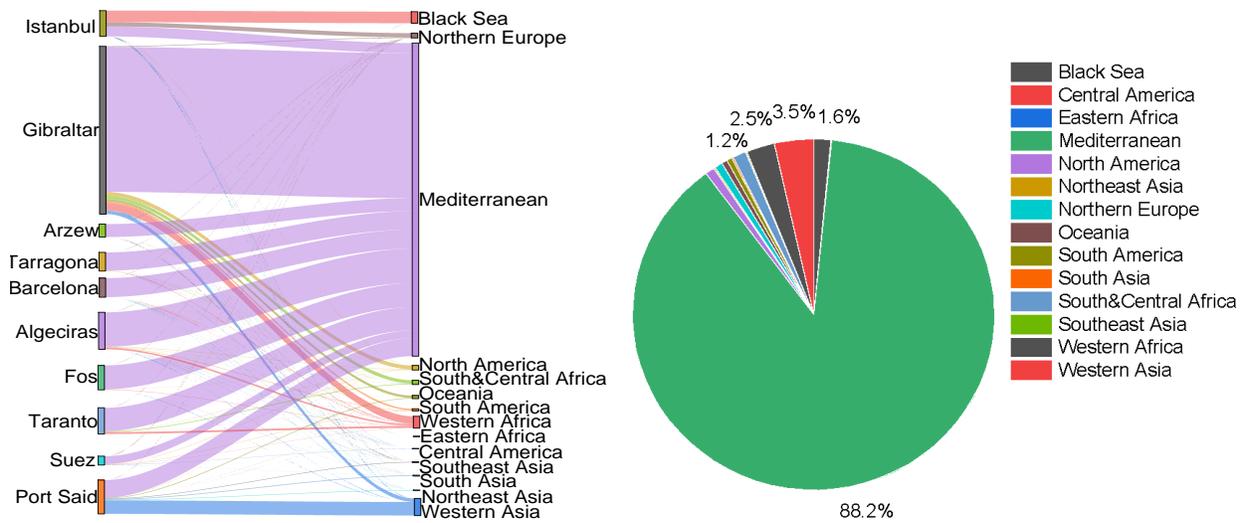

b. 2015

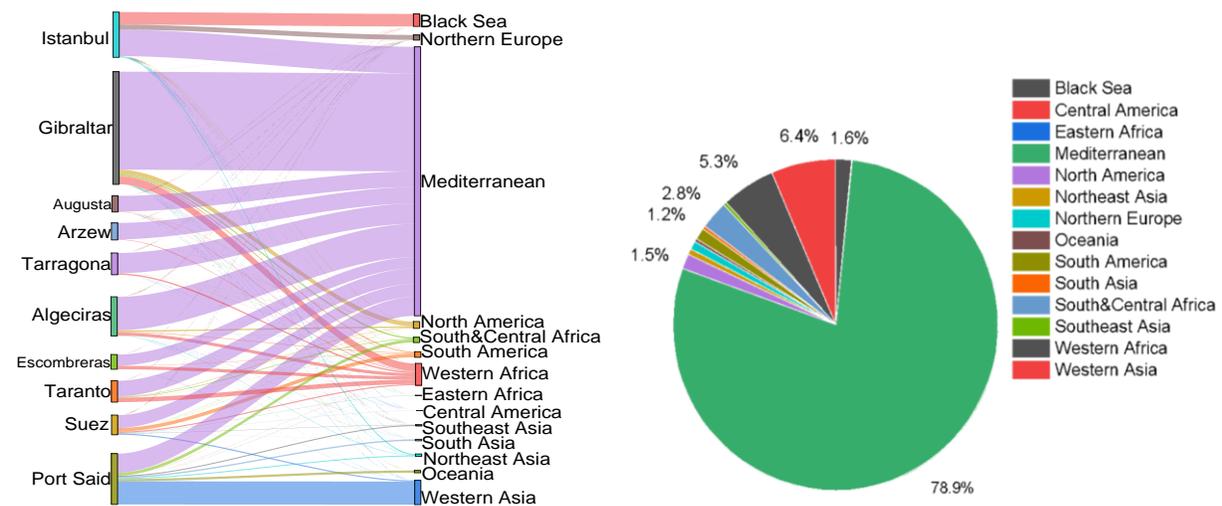

c. 2018

Figure 6 Species risk source regions in three years. Left panels: risk sources (right) to high-risk Mediterranean ports (left). Right panels: risk sources to the Mediterranean

### 3.4 Regulating high-risk ports connecting many clusters generates disproportionate risk reduction effectiveness

We found that when the treatment efficacy is 76%, the average risks of Mediterranean ports decrease by 9.4%, 11.0%, 12.4%, and 66.9% under Scenarios 1 to 4 by regulating 1, 2, 3, and 347 ports, respectively. It means under Scenarios 1 to 3 where high-risk ports are regulated, the average risk of Mediterranean ports can be reduced by 9.4%, 5.5%, and 4.1% by regulating one high-risk port; however, under Scenario 4, the average risk of Mediterranean ports can only be reduced by 0.2% by regulating one por (Table 2). This pattern is the same when the treatment efficacy increases to 99%. This proves that the practice of regulating the high-risk "hub ports" can disproportionately reduce the overall risk of the Mediterranean. The full list of port risks under different policy scenarios can be found in Data Availability.

Table 2 Regulating high-risk ports can disproportionately reduce the average risk of Mediterranean ports

|  |  | Efficacy=76% | | Efficacy=99% | |
|---|---|---|---|---|---|
| Policy scenario | Number of regulated high-risk ports | Average risk reduction of Mediterranean Ports | Average risk reduction by regulating one port | Average risk reduction of Mediterranean Ports | Average risk reduction by regulating one port |
| Scenario 1 | 1 | 9.4% | 9.4% | 11.8% | 11.8% |
| Scenario 2 | 2 | 11.0% | 5.5% | 13.9% | 6.9% |
| Scenario 3 | 3 | 12.4% | 4.1% | 15.5% | 5.2% |
| Scenario 4 | 347 | 66.9% | 0.2% | 85.0% | 0.2% |

Another interesting finding is that the risks of Gibraltar, Istanbul, and Suez have neglectable changes under the first three scenarios, even though they are the ports directly regulated. Under Scenario 4, the risk of Gibraltar decreases by 14.6% and 20.7%, while the risk reductions of Istanbul and Suez remain neglectable. See Table 3. The findings remain the same even though the regulation efficacy increases to 99%, meaning hub ports cannot be protected well only through increasing the regulatory standards. This is due to the very high traffic volumes to these ports, and the high traffic volumes that offset the ballast water treatment efficacy. This reveals that the regulation towards the previous calling ports, besides regulation the hub ports themselves, is necessary to reduce the risk of hub ports. The most efficient way is the global regulation, as revealed in Wang et al. (2021). This work proves the necessity of both global and regional regulation to reduce the risks of hub ports.

Table 3 Port risks and changes under four policy scenarios

|  | (1) Efficacy=76% | | | | (2) Efficacy=99% | | | |
|---|---|---|---|---|---|---|---|---|
| Port | Risk reduction under scenario 1 | Risk reduction under scenario 2 | Risk reduction under scenario 3 | Risk reduction under scenario 4 | Risk reduction under scenario 1 | Risk reduction under scenario 2 | Risk reduction under scenario 3 | Risk reduction under scenario 4 |
| Gibraltar | 0.0% | 0.0% | 0.0% | 14.6% | 0.0% | 0.0% | 0.1% | 20.7% |
| Istanbul | 0.1% | 2.1% | 2.1% | 1.1% | 0.1% | 2.8% | 2.8% | 2.4% |
| Suez | 0.0% | 0.0% | 0.0% | 0.0% | 0.0% | 0.0% | 0.0% | 1.8% |
| Algeciras | 0.0% | 0.7% | 0.8% | 42.0% | 0.0% | 0.8% | 0.9% | 58.3% |
| Port Said | 2.0% | 1.8% | 2.5% | 59.0% | 3.2% | 3.2% | 4.4% | 78.4% |
| Taranto | 30.0% | 30.6% | 30.6% | 64.2% | 36.8% | 37.7% | 37.7% | 78.7% |
| Escombreras | 0.0% | 0.0% | 0.0% | 55.2% | 0.0% | 0.0% | 0.0% | 72.0% |
| Arzew | 0.0% | 0.0% | 0.1% | 73.7% | 0.0% | 0.0% | 0.2% | 91.3% |
| Tarragona | 16.3% | 17.4% | 17.4% | 64.6% | 20.4% | 21.7% | 21.8% | 81.6% |
| Augusta | 3.9% | 3.9% | 6.3% | 72.1% | 4.5% | 4.5% | 7.0% | 90.5% |
| Barcelona | 27.2% | 27.2% | 27.2% | 63.4% | 33.5% | 33.5% | 33.5% | 80.5% |
| Nemrut Bay | 24.0% | 24.0% | 24.9% | 76.4% | 32.9% | 32.9% | 33.6% | 95.0% |
| Castellon | 23.1% | 23.8% | 23.8% | 69.2% | 26.9% | 27.8% | 27.9% | 86.0% |
| Fos | 45.9% | 46.6% | 46.6% | 65.8% | 54.2% | 55.1% | 55.1% | 79.3% |
| Sidi Kerir | 3.0% | 3.0% | 7.4% | 87.9% | 3.8% | 3.8% | 8.8% | 96.7% |

Note: the policy scenario is analyzed based on data of 2018. The risks in 2018 under different scenarios are normalized with regard to the risk of Gibraltar under no regulation for comparison.

## 4. Discussion and policy implications

Our work employs a higher-order network species spread risk model and clustering analysis to examine the port risks in the Mediterranean. We choose three years as the studied scope, 2012, 2015, and 2018, trying to consider the event of the Suez Canal expansion in 2015 and examine the potential impacts of this event on risk changes. We find the risks of most high-risk ports decreased from 2012 to 2018. With the examination of arriving traffic and ballast water discharge volume and analysis of higher-order patterns, we reveal the relationship between them and risks and analyze the reasons behind the risk changes.

No linear relationship between countries' risk change and traffic or BW change, except Gibraltar, whose risk change is positively related to its traffic and BW discharge volume. This shows the higher-order risk spread pattern plays an important role in determining the risk besides shipping traffic and BW discharge volume. At the individual port level, most ports experienced increased shipping traffic and BW discharge, yet decreased risks. This can be explained by two reasons. (1) Gibraltar has a determinant role in deciding species spread risks of other Mediterranean ports. See Figure 5, where the traffic and BW discharge changes of Gibraltar experienced a substantial drop in 2018. (2) The higher-order network plays an important role in decreasing the risk. For example, the number of connected clusters of high-risk ports decreased over time. This implies that the risk assessment needs to consider the high-order network pattern and geographical and environmental conditions.

"Strait ports", Gibraltar, Suez, Istanbul, and Algeciras, remain the high-risk ports in 2012, 2015, 2018, though the list of the top ten high-risk ports changed. Gibraltar remains in the top three ports, which connects the most clusters, followed by Suez and Port Said. These high-risk ports would be the key ports to regulate in the Mediterranean.

The highest species spread risk to the Mediterranean are from inside the Mediterranean due to the high traffic within this area and the closer location. The next source regions are Western Africa and the Black Sea due to the closer geographic locations. The species risk within the Mediterranean is smaller (from 90% in 2012 to 80% in 2018), while western Asia became more important in 2015 and 2018. This implies that the key to reduce species spread risk to the Mediterranean is to regulate the Mediterranean ports and closer located ports.

Gibraltar, Suez, Istanbul, and Algeciras remain the high-risk ports in 2012, 2015, 2018. These ports are also those connecting the most clusters. We hypothesized that regulating these high-risk hub ports may be more effective in reducing the overall risk in Mediterranean ports, so we designed and analyzed several policy scenarios to test this hypothesis. The policy scenario simulations reveal that by only regulating one high-risk port, the average Mediterranean risk can be reduced by 9.4%, 5.5%, and 4.1% while regulating one average-risk port will only reduce risk by 0.2% across all Mediterranean ports when the treatment efficacy is 76%. The average Mediterranean risk can be reduced by 11.8%, 6.9%, and 5.2% while regulating one average-risk port will only reduce risk by 0.2% across all Mediterranean ports when the treatment efficacy is 99%. This proves that the practice of regulating several major high-risk "hub ports" can disproportionately reduce the overall risk of the Mediterranean.

Another interesting finding is that the risks of Gibraltar, Istanbul, and Suez have neglectable changes when these ports are directly regulated. This reveals the determining role of arriving shipping traffic, which offsets the ballast water treatment efficacy.

Therefore, we obtain different policy implications for reducing risks to the Mediterranean and reducing risks of high-risk ports. To reduce the overall risk to the Mediterranean, hub ports are very important to control to generate a disproportionate risk reduction effectiveness. To reduce the risk of single high-risk hub ports (high-risk ports tend to be hub ports and vice versa), the regulation towards themselves is not adequate and needs regulation towards previous calling ports.

One result is different from the work of Sardain et al (Sardain et al., 2019). Their results show species from far regions may have higher risks, while our results show higher risks tend to happen to closer located regions. While we do agree that places far away from each other may have species invasion due to the ballast water carriage all over the world. We do observe risks from Asia and America to the Mediterranean, as proved by the literature. For example, Shefer et al. claim that haplotypes at any site may include those introduced from remote sources, hence reducing the correlation between genetic and geographical distance due to human-mediated introductions (Shefer et al., 2004). However, Sardain et al. found Northeast Asia is the biggest source of species spread to the Mediterranean. Besides the difference caused by our higher-order network, another important reason is the difference in the modeling of the likelihood of being nonindigenous (the first item of Equation 1). We assume the nonindigenous probability between two ports if 1 if the ports do not belong to the same or neighboring ecoregions (Costello et al., 2017, Abell et al., 2008), while Sardain et al. use a function of biogeographical dissimilarity, which depends on inter-port distance. In our method, there are nine different ecoregions in the Mediterranean (Spalding et al., 2007) and the values of nonindigenous probability of many port pairs are 1. In the work of Sardain et al, these ecoregions are geographically close to each other, making the probability much smaller (close to 0, rather than 1). This reveals that different assumptions of P(alien)'s effect is dominant in determining the spread risk. The substantial difference in findings by using different species spread risk models shows the necessity of robustness analysis when making policy suggestions based on risk model-based scientific evidence.

Though the higher-order network risk spread risk model has been used in past work (Saebi et al., 2019, Saebi et al., 2020a). Limitations need to be noticed. Also, we do not consider environmental data change from 2012 to 2018 to better examine the impact of shipping traffic changes. Rising temperatures and other climate-driven changes have already shifted geographic distributions of marine species by boosting thermophilic NIS (Occhipinti-Ambrogi, 2007, Van der Putten et al., 2010). If the process of global warming continues, the risk that additional warm-water species become established in the Baltic Sea will increase (Boxshall et al., 2007).

The species invasion risk to the Mediterranean comes from several vectors (Zenetos et al., 2012) and we only describe the introduction from ballast water discharge of ships in this work. The vector of the Suez Canal is different from the introduction via shipping traffic through the Suez Canal. We focus on the introduction to Mediterranean ports, including the introduction to Suez port, so the research on introductions via the Suez Canal is not within our scope.

**Data availability**

Ship movement data from Lloyd's List Intelligence are purchased and are not publicly available. Other data obtained are publicly available and cited. The full lists of port risks in three years, clusters of all Mediterranean ports, and policy simulation results are available at https://github.com/msaebi1993/Species-Flow-Networks/tree/master/data/MedSea.

**Code availability**

The Code for generating higher-order network and synthetic data is available at https://github.com/msaebi1993/HON-ANOMALY.

**Acknowledgement**

The work is based on the NSF Coastal SEES project [grant number 1748389]. We thank NSF Coastal SEES for funding this work.